\documentclass[10pt]{article}
\usepackage[OE]{express}
\usepackage{color}
\usepackage{float}
\usepackage{graphicx}
\usepackage{subfig}
\begin{document}
\title{Binarized Gerchberg Saxton Algorithm for Hologram Generation Using a Digital Micromirror Device}

\author{Chunde Huang\authormark{1} }

\address{\authormark{1}Physics Department of Washington State University\\
}

\email{\authormark{*}chunde.huang@wsu.edu} 



\begin{abstract}
In this paper, a modified Gerchberg Saxton algorithm for generating improved robust binary hologram is presented.
\end{abstract}

\section{Introduction}

A laser can introduce atom-light interaction via dipole interaction. In cold atom physic, lasers are widely used to manipulate atoms, such as Rabi flopping, dipole trapping, optical lattices, potential barrier and phase mask, etc. It's also of great interest to construct quantum gas microscope \cite{bakr2009quantum} for single-site addressability, which provides a very efficient method for manipulating the quantum state of a cold atom. When many such sites form a lattice, those trapped atoms can be used for quantum computing. Such an application requires a very high precision of lattice generation, which is hard when there are all kinds of aberration in an optical setup. To undo the wavefront distortion, one method is to use the binary hologram to compensate such imperfection \cite{zupancic2016ultra, zupancic2013dynamic}.

In cold atom experiments, we also want to trap a BEC in all kind of dipole potential to study its quantum dynamic. Besides, we also want to modify the trap in real-time and observe how a BEC evolves. However, It's very challenging to generate arbitrary dipole potential profiles with the desired phase map using transitional methods, such as transparency. The advent of the programmable digital micro-mirror device(DMD) gives scientists new freedom to create arbitrary dipole potential\cite{bowman2014red, gaunt2011robust, gauthier2016configurable, ren2015dynamic}.

\section{Digital Micromirror Device}
\begin{figure}[htb]
	\begin{center}
		\includegraphics[scale=0.35]{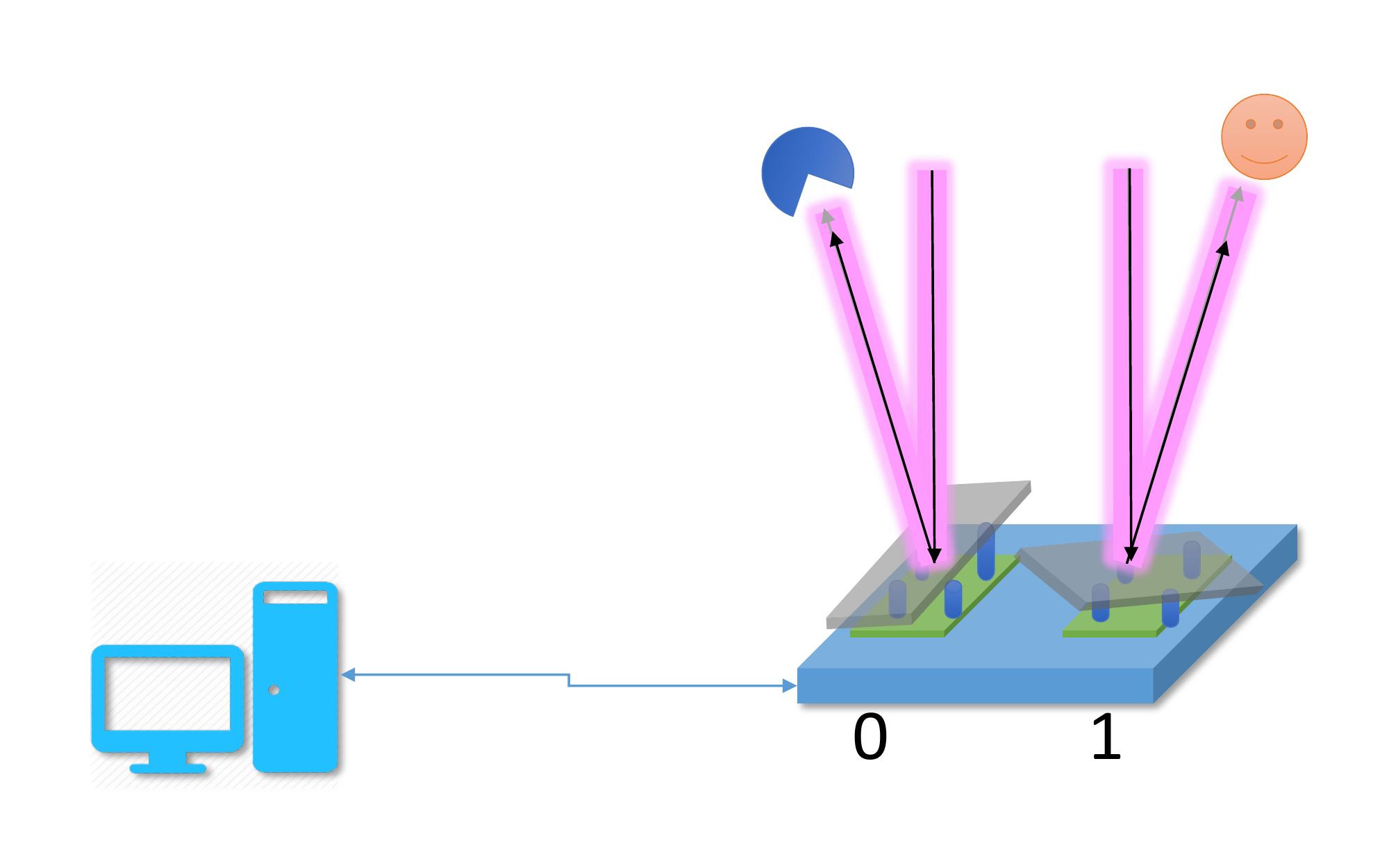}
	\end{center}
	\caption[Different State of Micromirrors]{Digital Micromirror Device Diagram of Two Mirrormirrors with Different States.}
	\label{fig:dmd_diagram}
\end{figure}

\begin{figure}[htb]
	\begin{center}
		\includegraphics[scale=0.35]{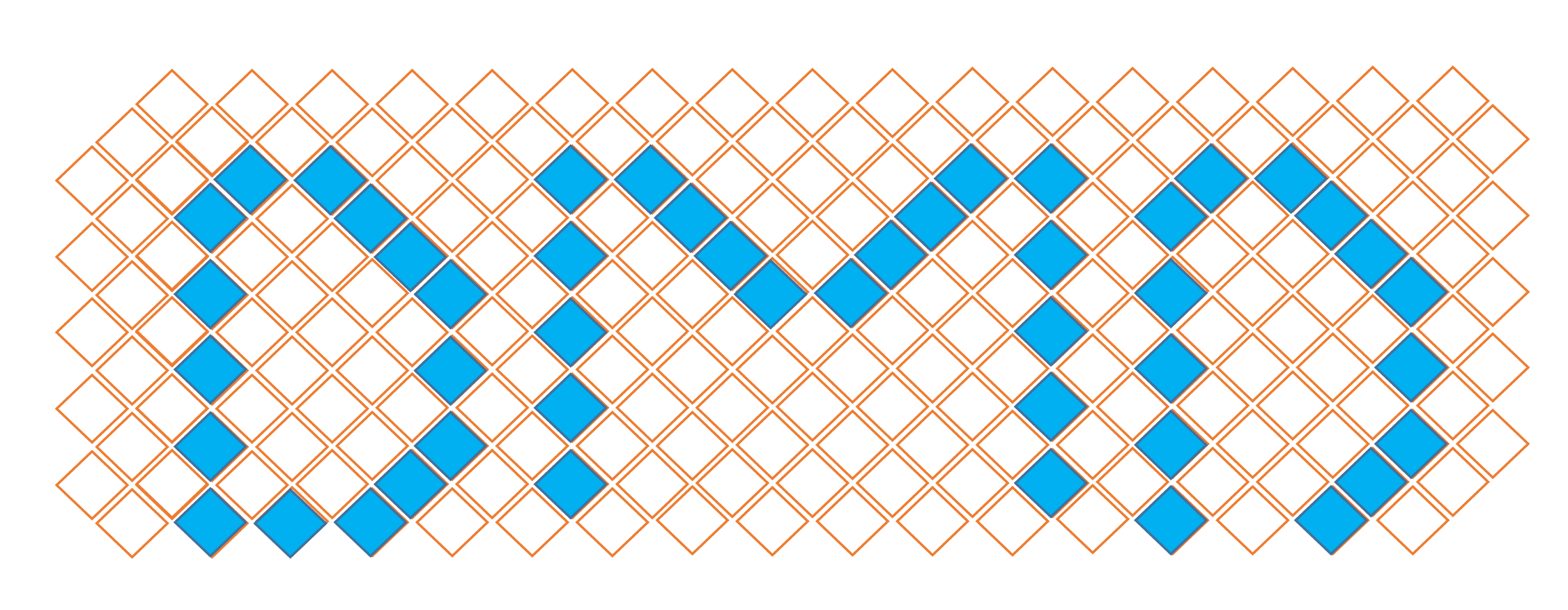}
	\end{center}
	\caption[Pattern on a DMD]{One example of patterns a DMD. The filled diamonds represent mirrors of state on, while those empty diamonds represents mirrors with state off.}
	\label{fig:dmd_pattern}
\end{figure}

A Digital Micromirror Device(DMD) contains a 2D array of very tiny mirrors. Each mirror has a control memory bit associated with it, that enables the mirror to be turned on or off independently. When a mirror is on, it will reflect light to the desired direction, when it is off, the light will be reflected in another direction. A diagram of a DMD is shown in the Fig.\ref{fig:dmd_diagram}.  Such a device is connected to a computer that controls the state of each micromirror. A DMD contains large mount of such tiny mirrors, and we are able to control the pattern of a DMD formed by its micromirror array. An example of patterns on a DMD is showed in the Fig.\ref{fig:dmd_pattern}, in which each of those micromirror is rotated by 45 degree to have a shape of a diamond. Some other types of DMD have different layout of micromirror. 

\begin{figure}[htb]
	\begin{center}
		\includegraphics[scale=0.45]{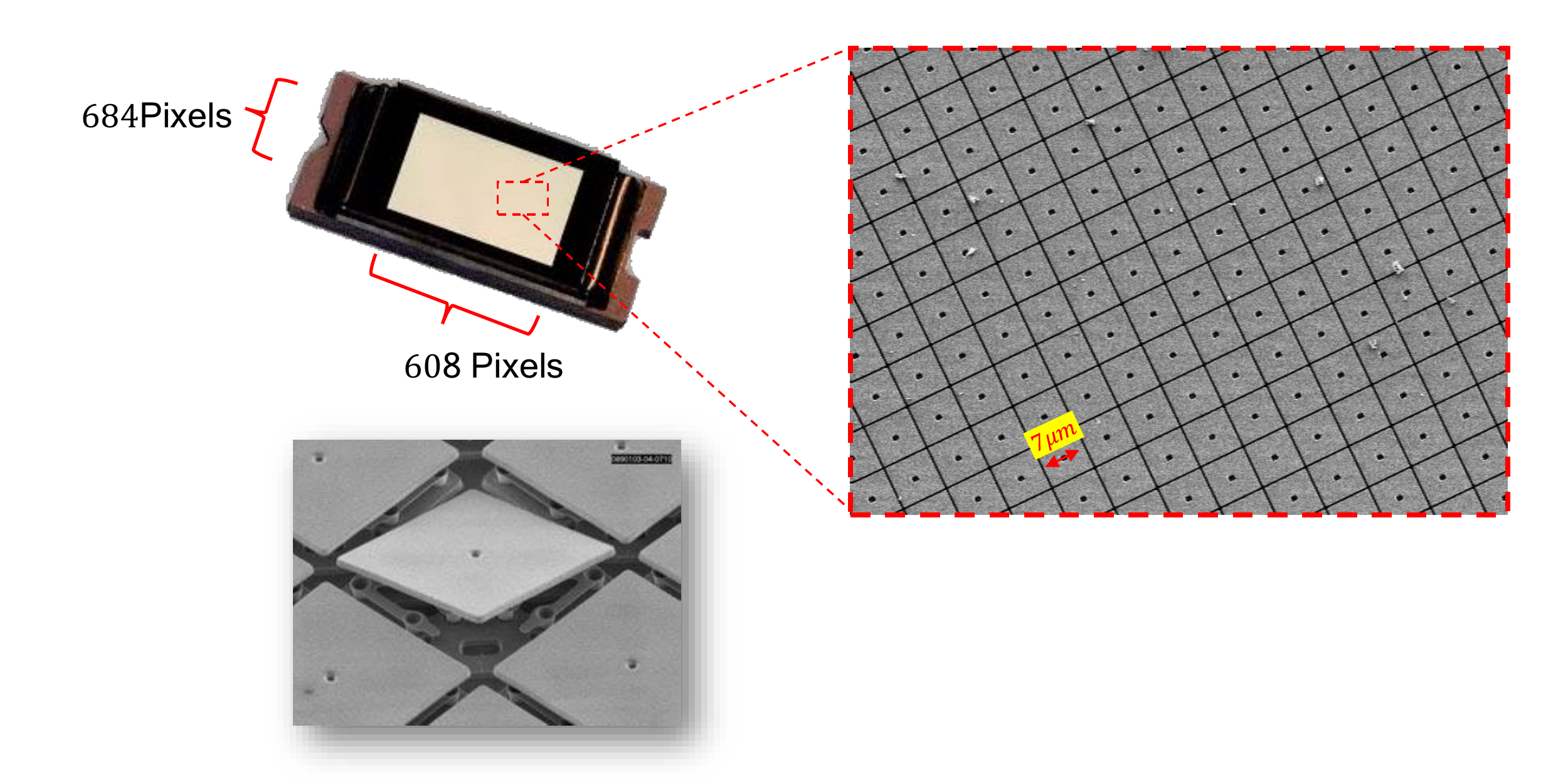}
	\end{center}
	\caption[Physical Geometry of a DMD]{This is a toy model of DMD used for testing the idea of spatial light modulator Its micromirror array has dimension 608$\times$684 in unit of pixels, each pixel is a micromirror}
	\label{fig:dmd_geometry}
\end{figure}

The physical geometry of a real DMD(Model DLP3000 from Texas Instruments) is shown in the Fig.\ref{fig:dmd_geometry}. This is the first type of DMD we used for experimental tests. The size of each micromirror is about 7$\mu m$, and can be turned on and off around 4000Hz. Some higher models can provide a much higher resolution and refresh rate. 

\section{Fourier Imaging}
\begin{figure}[H]
	\begin{center}
		\includegraphics[scale=0.6]{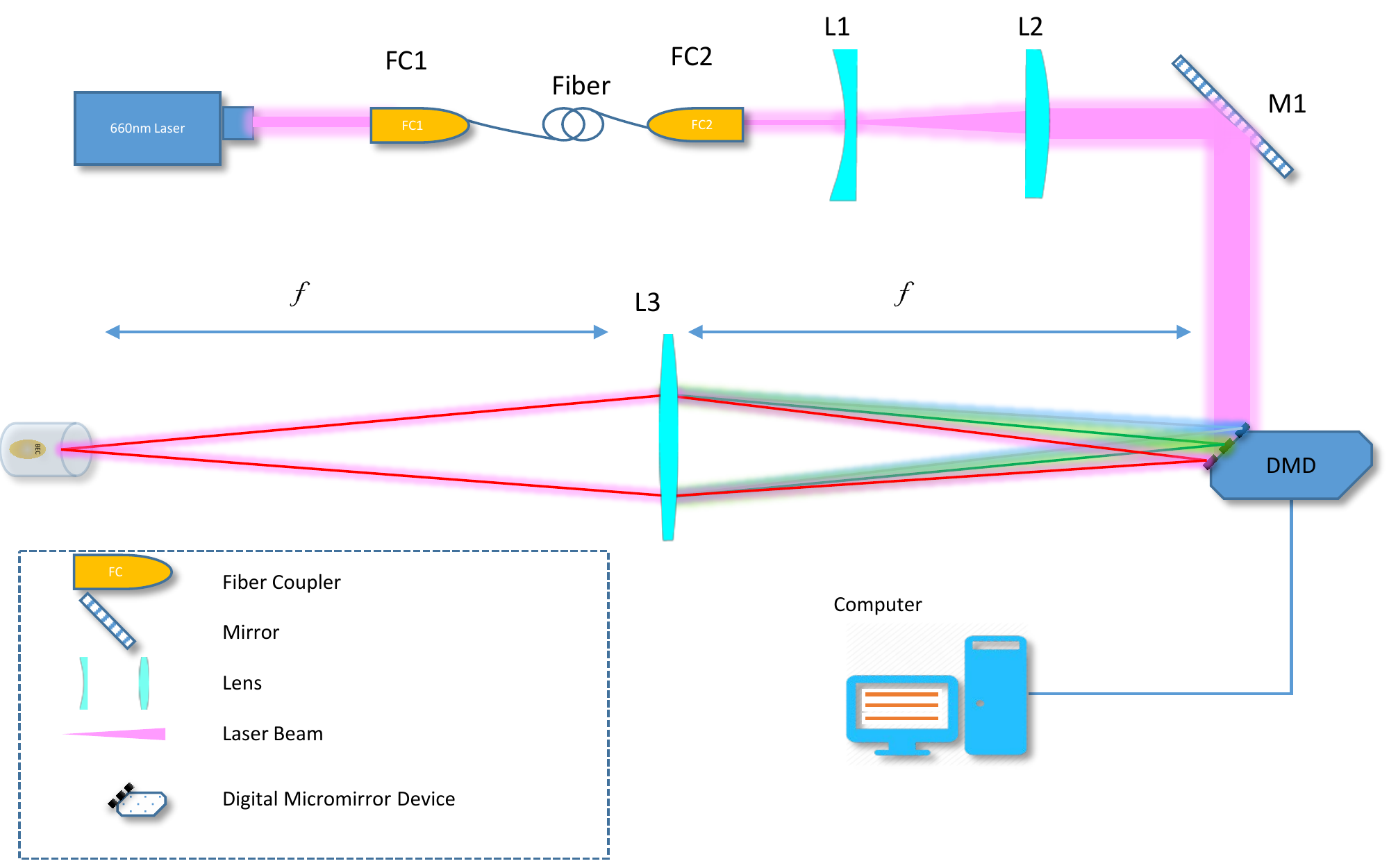}
	\end{center}
	\caption[Optical Setup for Fourier Imaging]{Optical System Setup for Fourier Imaging. The DMD and the target planes are all at the focusing length of the lens L3. L1 and L2 act ass telescope to expand the laser beam.}
	\label{fig:DMD_Fourier_Geometry}
\end{figure}

\begin{figure}[H]
	\begin{center}
		\includegraphics[scale=0.6]{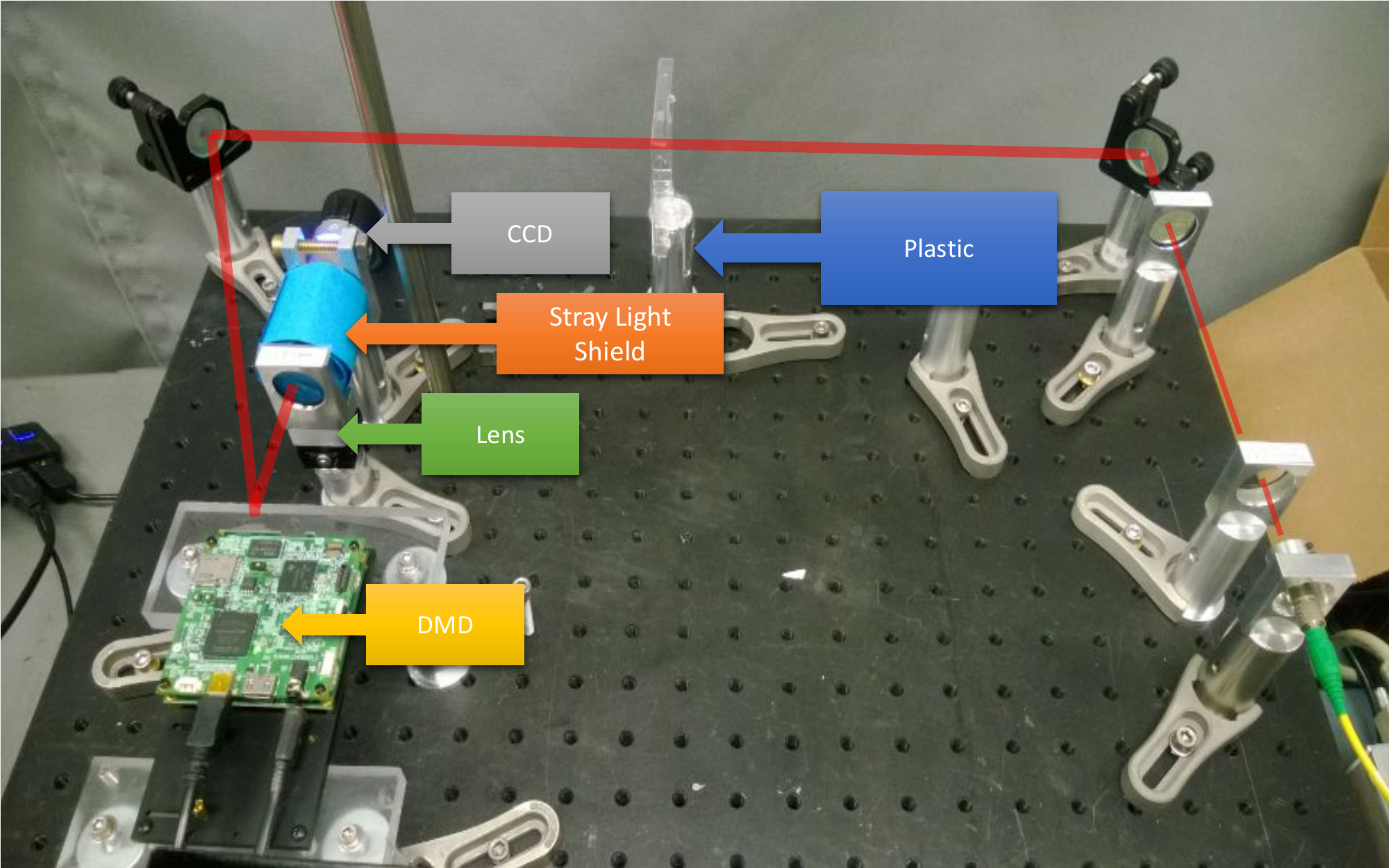}
	\end{center}
	\caption[Actual Optical Setup for Fourier Imaging]{Actual Optical Setup for Fourier Imaging. The CCD and DMD are located at the focusing lengths of the lens, a blue tube is used for shielding the stray light to increase the contrast. A plastic slide is used to introduce more distortion to the wavefront.}
	\label{fig:DMD_Fourier_Real_Setup}
\end{figure}

A thin lens can be used as a Fourier engineer to transform a DMD pattern to the target image, the efficiency of light focused on the target plane is not necessarily lower than the case of direct imaging, but it offers the possibility to modulate the phase profile of the target. The optical system setup is shown in the Fig.~\ref{fig:DMD_Fourier_Geometry}. The actual optical setup is shown in the Fig. ~\ref{fig:DMD_Fourier_Real_Setup}.

\section{Binarized Gerchberg-Saxton Algorithm}

The algorithm described in \cite{zupancic2013dynamic}) gives the general idea of generating binary hologram, which yields very good results. But it needs to use some dithering or other methods to improve the image quality, which may not necessarily yield the best result. In the thesis, a modified version of Gerchberg-Saxton(GS) algorithm is proposed, it takes the binary nature of the DMD into account. The conventional GS algorithm can be done in two different ways, one of them is shown in Table \ref{table:General_GS_Algorithm}

\begin{table}[H]
	\centering
	\begin{tabular}{|p{13cm}|}
		\hline
		
		\textbf{Algorithm: Gerchberg-Saxton Algorithm}\\
		1, $A = IDFT(T)$, where T is the target image, $IDFT(x)$ is the inverse  Fourier transform.\\
		2, $B= Amp(S) \times e^{i \phi(A)}$, where S is the source image, $Amp(X)$ is the function returns the amplitude of a complex number $x$, $\phi(x)$ returns the phase of $x$.\\
		3, $C= DFT(B)$, where $DFT$ is the Fourier transform.\\
		4, $D=Amp(T)\times e^{i \phi(C)}$\\
		5, $A= IDFT(D)$.\\
		6, Check if the change of $A$ is converged. if not, return to step 2.\\
		\hline
	\end{tabular}
	\caption{Algorithm: Gerchberg-Saxton Algorithm}
	\label{table:General_GS_Algorithm}
\end{table}

\begin{figure}[H]
	\begin{center}
		\includegraphics[scale=0.7]{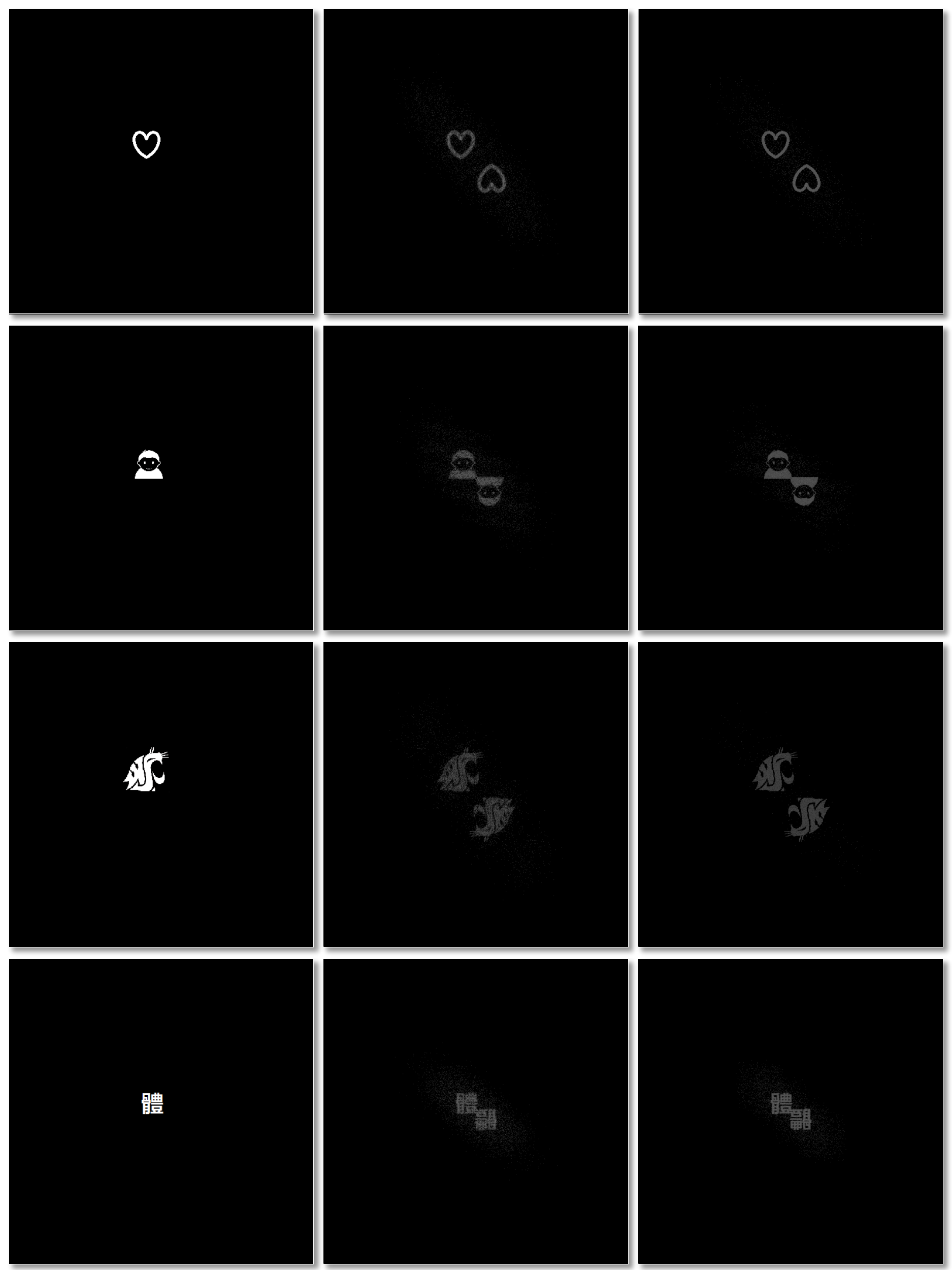}
	\end{center}
	\caption[Comparison of GS algoritms]{All results are with 6 iterations. The first column is the target images, the middle column is the results from conventional GS algorithm, while the third column is the results using the binarized GS algorithm. Zoom in to check the detail for an on-line copy.}
	\label{fig:GS_Comparision}
\end{figure}

\begin{figure}[H]
	\begin{center}
		\includegraphics[scale=0.5]{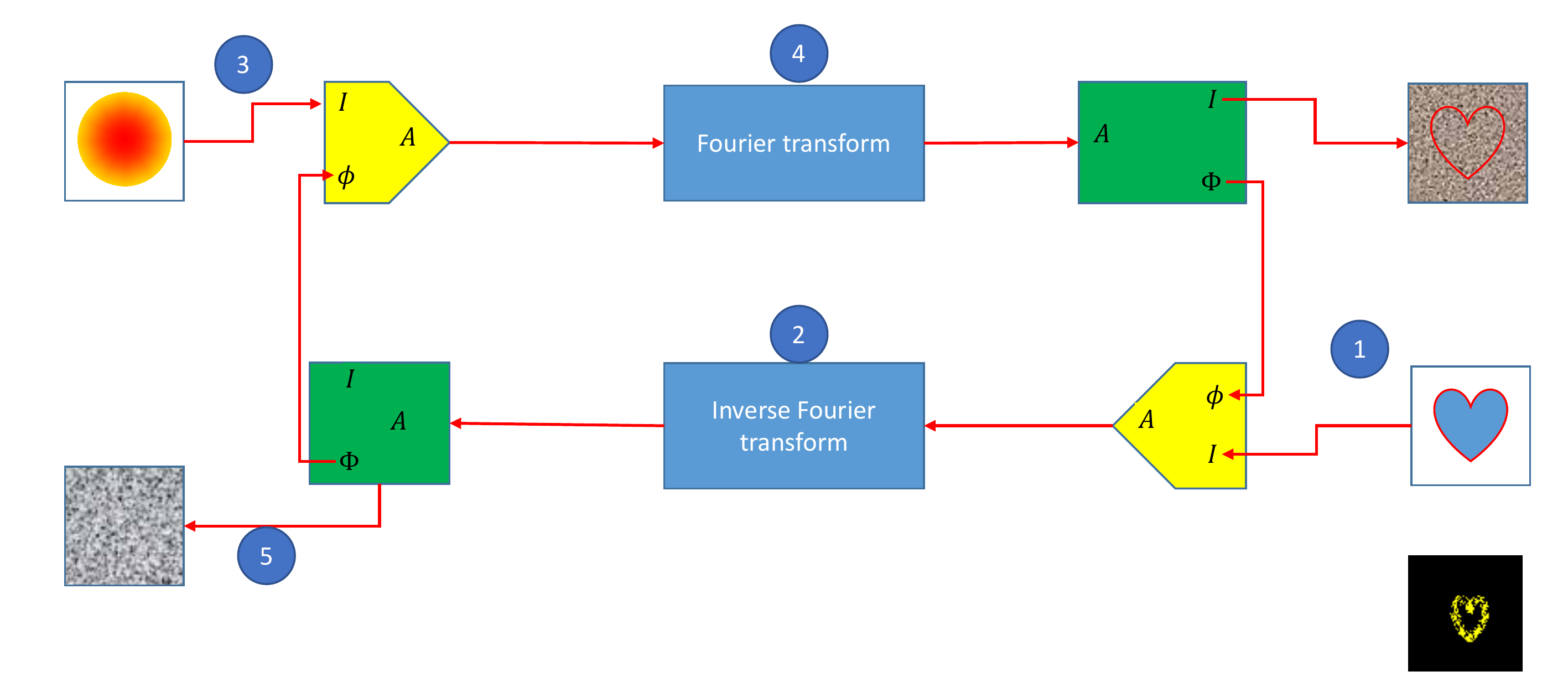}
	\end{center}
	\caption[Gerchberg Saxton Algorithm]{Gerchberg Saxton Algorithm: The blue box is either Fourier transform or its inverse operation. The yellow pentagon represents the combination of amplitude and phase input to generate a complex map as output. The green box is taking complex as input, and output two real components(Amplitude and phase). Step 5 output the hologram phase map. Step 3 takes the laser profile as input. Step one takes the desired target image as input.}
	\label{fig:GS_Algorithm}
\end{figure}

\begin{figure}[H]
	\begin{center}
		\includegraphics[scale=0.5]{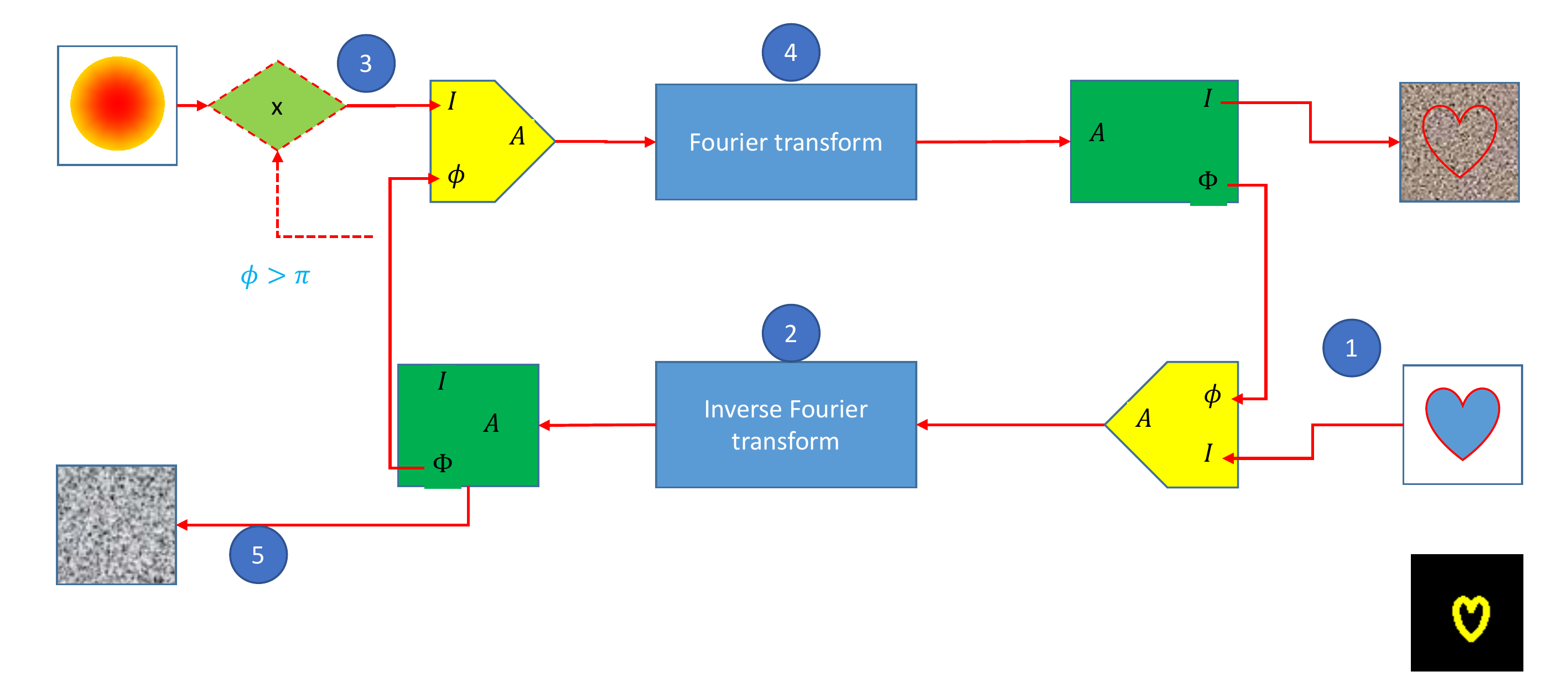}
	\end{center}
	\caption[Binarized Gerchberg Saxton Algorithm]{Binarized Gerchberg Saxton Algorithm: All the steps are the same as the GS algorithm shown in Figure \ref{fig:GS_Algorithm}. The only difference comes from the diamond shape part when the laser profile is changed based on the phase of step 2, if the phase of a pixel is smaller than $\pi$, the intensity of that point input at step 3 will be set to zero.}
	\label{fig:BGS_Algorithm}
\end{figure}

The GS algorithm can also be presented as a flow chart as shown in Fig. \ref{fig:GS_Algorithm}. The problem to apply this directly to the DMD binary hologram calculation is obvious, the final result has to be binarized using some cut-off value, which can be too coarse. If a simple rule of binarization is used, for example, all resulted value with phase less than $\pi$ is set to one, otherwise zero, the image quality is not good, as can be seen in the Fig. \ref{fig:GS_Algorithm} and Fig. \ref{fig:GS_Comparision}.
To address this issue, one more step is inserted into the GS algorithm, as shown in the Fig.\ref{fig:BGS_Algorithm}, we term the modified GS algorithm "Binarized GS Algorithm" or BGS algorithm. The simple one-step makes a big difference, the resulted image from the simulation is much better. \\
So far, the distorted phase map has not been taken into the calculation, in real experimental, the phase correction is done by adding the phase map to the output of step 5, and then binarizing the result using a simple rule, i.e: set the value of a pixel to one of the corresponding phases is smaller than $\pi$, and zero otherwise. The image quality of the BGS method in much better than the GS method, and it's also better than the method mentioned in \cite{zupancic2013dynamic}. 

\begin{table}[H]
	\centering
	\begin{tabular}{|p{13cm}|}
		\hline
		\textbf{Algorithm: Binarized Gerchberg-Saxton Algorithm}\\
		1, $A = IDFT(T)$, where T is the target image, $IDFT(x)$ is the inverse  Fourier transform.\\
		2, $B= Amp(S) \times \theta (\phi(A))$, where S is the source image, $Amp(X)$ is the function returns the amplitude of a complex number $x$, $\phi (x)$ returns the phase of $x$. $\theta (x)$ is a binary function which returns 1 when $x < \pi$, 0 otherwise.\\
		3, $C= DFT(B)$, where $DFT$ is the Fourier transform.\\
		4, $D=Amp(T)\times  e^{i \phi(C)}$\\
		5, $A= IDFT(D)$.\\
		6, Check if the change of $A$ is converged. if not, return to step 2.\\
		\hline
	\end{tabular}
	\caption{Algorithm: Gerchberg-Saxton Algorithm}
	\label{table:Binarized_General_GS_Algorithm}
\end{table}

The improvement from the binarized GS algorithm is shown in the Fig. \ref{fig:GS_Comparision}, and it can be seen that the modified GS algorithm gives more accurate images, and also with less surrounding noise for a few iterations. When the iteration number increases, the resulted quality of the GS algorithm will improve and will close to that produced by the BGS algorithm. So the major improvement of the BGS over GS is the speed of convergence.

\begin{figure}[H]
	\begin{center}
		\includegraphics[scale=0.53]{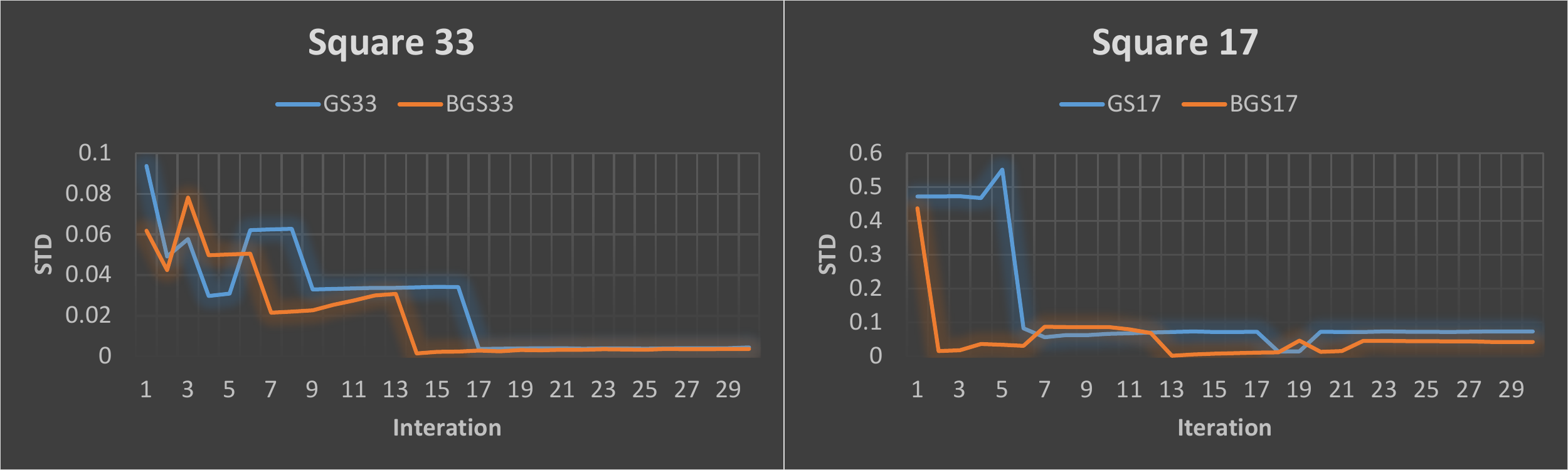}
	\end{center}
	\caption[Standard Deviation of Pixel Values]{Standard deviation of pixel values at the first order of resulted images as iteration increases. the GS33 line is for the case for a uniform square with a size 33 pixel produced by the GS algorithm. The BGS33 is for the BGS algorithm. The square 17 plot is for a uniform square wit size 17 pixels. The maximum iteration is 30, it can be seen that the BGS converges faster than the GS, especially for the case of the small square, it converges in just about 3 iterations, while the GS take more than 6 iterations to converge.}
	\label{fig:Pixel_Value_STD}
\end{figure}

\section{Conclusion}
In this paper, a modified version of Gerchberg Saxton algorithms by taking the binarization procedure of generating hologram for a DMD into consideration is presented. By simulation, the improvement of convergence is verified.
 
\bibliographystyle{unsrt}
\bibliography{BinarizedGerchbergSaxtonAlgorithm}
\end{document}